\documentclass{bcp}

\usepackage{amsmath,amssymb,amsfonts}
\usepackage{graphics,graphicx,epsfig,pstcol,psfrag}

\usepackage{mathbbol}
\usepackage{mathrsfs}
\usepackage{bbm}
\usepackage{bbold}

\def\LaTeX{L\kern -.36em\raise .3ex\hbox{\sc a}\kern -.15em T\kern -.1667em%
\lower .7ex\hbox{E}\kern -.125em X}

\def\be{\begin{equation} \label}
\def\ee{\end{equation}}

\def\EE{\mathbb{E}}

\def\ZZ{\mathbb{Z}}

\def\A{\boldsymbol{A}}
\def\G{\boldsymbol{G}}

\def\R{\boldsymbol{R}}
\def\U{\boldsymbol{U}}

\def\c{\boldsymbol{c}}

\def\m{\boldsymbol{m}}
\def\e{\boldsymbol{e}}
\def\n{\boldsymbol{n}}

\def\X{\mathcal{X}}

\newcommand{\1}{\ensuremath{\mathbb{1}}}

\def\ti{\to\infty}
\renewcommand{\sp}[1]{\langle #1\rangle}

\begin{document}
\bibliographystyle{abbrv}

\renewcommand{\figurename}{Figure}{\small Figure}

\mathclass{Primary 92D15; %Problems related to evolution
Secondary 60J70.} %Applications of diffusion theory
\thanks{}
\abbrevauthors{E.~Baake and R.~Bialowons}
\abbrevtitle{Ancestral processes  with selection}

\title{Ancestral processes  with selection: \\ Branching and Moran models}

\author{Ellen Baake}
\address{Technische Fakult\"at, Universit\"at Bielefeld\\
Postfach 100131, 33501 Bielefeld, Germany\\
E-mail: ebaake@techfak.uni-bielefeld.de}

\author{Robert Bialowons}
\address{Institut f\"ur Mathematik und Informatik, Universit\"at Greifswald\\
Jahnstr. 15a, 17487 Greifswald, Germany \\
E-mail: robert.bialowons@uni-greifswald.de}

\maketitlebcp

\abstract{
We consider two versions of stochastic population models with mutation
and selection.  The first approach relies on a multitype branching
process; here, individuals reproduce and change type (i.e., mutate)
independently of each
other, without restriction on population size.  We analyse the
equilibrium behaviour of this model, both in the forward and in the
backward direction of time; the backward point of view emerges if the
ancestry of individuals chosen randomly from the present population is
traced back into the past. 

The second approach is the Moran model with selection. Here, the
population has constant size $N$.  Individuals reproduce (at rates
depending on their types), the offspring inherits the parent's type,
and replaces a randomly chosen individual (to keep population size
constant). Independently of the reproduction process, individuals can
change type. As in the branching model, we consider the
ancestral lines of single individuals chosen from the equilibrium
population.  We use analytical results of Fearnhead (2002) to
determine the explicit properties, and parameter dependence, of the
ancestral distribution of types, and its relationship with the
stationary distribution in forward time.
}

\section{Introduction.}
\label{sec:intro}
Like most dynamical processes, stochastic models of biological populations
are usually considered in the forward direction of time. Ancestral 
processes arise if attention is shifted to the past history of a population,
given its present state. This backward point of view is
particularly important in the area of {\em population genetics}, which
is concerned with the genetic structure of populations under the
joint action of evolutionary forces like mutation and selection.
Here, the large time scales involved usually make it impossible
to observe the dynamics over any relevant time span;
one therefore resorts to taking samples from present-day populations,
and inferring their history, in a probabilistic or statistical sense.
More precisely, starting from  {\em population sequence data}
(sequences from a single locus analysed in a sample of individuals
of the same species), three different sorts of inference questions can
be addressed (see \cite{StDo03}):

\begin{enumerate}
 \item  What are the genetic parameters (mutation rates, selective 
       intensities) governing the dynamics of the process?
 \item The past history of the genetic types observed in the
       sample; for example, the age of a mutation (e.g., when has a
       disease gene arisen?)
 \item The demographic history of the population from which the
       data was sampled; for example, when did the most recent
       common ancestor live?
\end{enumerate}

For an overview of the area and the corresponding inference
techniques, see \cite{StDo03}. Clearly, 
such endeavour relies crucially on a good knowledge of the population process
viewed in the reverse direction of time.

We will discuss such ancestral processes here for two classes of models 
that describe the interaction of mutation and selection (or, more
generally, type-dependent reproduction). We will first consider
certain (multi-type) branching processes, where individuals mutate and
reproduce independently of each other, and, in particular,
independently of the current population size, which is, therefore, free
to fluctuate. At the other extreme, we will consider the so-called
Moran model, which assumes that birth events are strictly coupled to
death events (any newborn individual replaces one of the
individuals already present), so that the population size remains constant
throughout. 

The purpose of the article is twofold. The larger part of the paper
will explain, in a tutorial
manner, the forward and the ancestral processes in both types of models,
with special emphasis on the asymptotic behaviour. 
We will start with the branching process 
(Sec.~\ref{sec:branching}--\ref{sec:twotype}), because it is simpler 
due to the independence
of individual lines. The corresponding backward process
is analogous to the usual time reversal of a Markov chain, and
directly leads to the composition of that part of the 
population that will become ancestral to future generations.
Genealogies (of individuals sampled from today's population)
are not relevant, and not required, in this context.

In contrast, reversing time in the Moran model is more intricate,
but also yields more information. Genealogies of individuals
sampled today play a central role here. As long as selection is
absent (i.e., all individuals reproduce at the same rate), genealogies
are easily constructed through the famous Kingman's coalescent
\cite{King82a,King82b}, which starts with a sample from today's
population and proceeds backward, joining pairs of lines at
certain rates and thus producing genealogical trees. This process
is quite tractable due to the crucial fact that the lineages of the sampled
individuals may be considered independently of the remaining population.
The properties of the resulting genealogies are, therefore, quite 
well understood.

This lucky situation changes when individuals reproduce at different
rates. Due to the coupling of birth and death events, individual
lineages now cease to be independent of the (type composition of)
the remaining population, and a more intricate construction must be
used, which is known as the \emph{ancestral selection graph}.
This is no longer a tree, but a branching/coalescing graph, into
which all possible genealogies are embedded, and the `true' one
(in the porbabilistic sense) can be extracted according to
certain rules. We will put forward the Moran model, and its
backward constructions, in Secs.~\ref{sec:moran}--\ref{sec:stationary}.

So far, our exposition will mainly
be a review of results available in the literature, but not easily
accessible for the uninitiated. In Secs.~\ref{sec:example}
and \ref{sec:virtuals}, we will then  present
some new results on a two-type example with asymmetric mutation,
which illustrates the presented concepts, and establishes
connections between the Moran and the corresponding branching models.

\section{Multitype branching: the mutation-\-reproduction model.}
\label{sec:branching}

Let $S$ be a finite set of types (with $|S|>1$), and consider a population 
of individuals, each characterised by
a type $i \in S$. We will think of types as {\em genotypes}, since children inherit
their parents' type. 
(Genetically speaking, our individuals are haploid, i.e., carry only one
copy of the genetic information per cell,
and  types are alleles.)
\begin{figure}[h]
  \centerline{\input{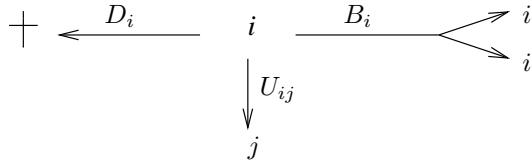}}
  \caption{\label{fig:museP} {\small The parallel mutation-reproduction model.}}
\end{figure}

We consider the most basic mutation-reproduction model, in which
mutation and reproduction  
occur in parallel.
As depicted in Fig.~\ref{fig:museP},
an individual of type $i \in S$ may, at any instant in continuous time, 
do either of three things: It may split, i.e., produce a copy
of itself (this happens 
at birth rate $B_i \geq 0$), it may die (at rate $D_i \geq 0$), 
or it may mutate to 
type $j,j \neq i$ (at rate $U_{ij} \geq 0$). This defines a
\emph{multi-type Markov branching process in continuous time} (see
\cite[Ch.\ V.7]{AtNe72}  or \cite[Ch.\ 8]{KaTa75}
for a general overview, and \cite{BaGe06} for the context
considered here). That is, 
an $i$-individual waits for an exponential time with parameter $A_i =B_i+D_i + 
\sum_{j:j\neq i} U_{ij}$, and then dies, splits or mutates to type $j \neq i$ with 
probabilities $B_i/A_i, D_i/A_i$, and $U_{ij}/A_i$, respectively.
The number of 
individuals of type $j$ at time~$t$, 
$Z_j(t) \in \ZZ_{\geq 0}:=\{ 0,1,2 \dots \}$, 
is a random variable; the collection  $Z(t)=\big(Z_j(t)\big)_{j \in S}$ is a random vector.
The corresponding expectation is described by the 
first-moment generator $\A=\U+\R$. Here, $\U$ is the Markov generator 
$\U=(U_{ij})_{i,j\in S}$, where the mutation rates $U_{ij}$ for $j \neq i$
are complemented by
$U_{ii}:= - \sum_{j:j\neq i} U_{ij}$ for all $i \in S$. 
Further, $\R:= \mbox{diag}\{ R_i 
\mid i \in S\}$, where $R_i:= B_i - D_i$ is the net reproduction rate 
(or {\em Malthusian fitness}) of an $i$-individual. 
More precisely, we have $\mathbb{E}^i (Z_j(t)) = 
(e^{t\A})_{ij}$, where $\mathbb{E}^i (Z_j(t))$ is the expected number 
of $j$-individuals at time $t$ in a population started by a single 
$i$-individual at time $0$.

\section{Properties of the branching model.}
\label{sec:properties}

We will assume throughout that $\A$ (or, equivalently, $\U$) 
is irreducible. Perron-Frobenius 
theory then tells us that 
$\A$ has a principal eigenvalue $\lambda$ (namely a real eigenvalue 
exceeding the real parts of all other eigenvalues) and associated 
positive left and right eigenvectors $\pi$ and $h$, which will be 
normalised so that $\langle\pi,\1\rangle = 1 = \langle \pi, h \rangle$, 
where $\1 = (1)_{i\in S}$ denotes the vector with all coordinates equal to $1$,
and $\sp{. \, ,.}$ is the scalar product.
We will further  assume  that $\lambda >0$,
i.e., the branching process is \emph{supercritical}. This implies that 
the population will, in expectation,  grow in the long run; in individual 
realisations,
it will survive with positive probability, and then grow to
infinite size with probability one, see \eqref{eq:growthrate} below. 
The asymptotic properties of our model  are, to a large 
extent, determined by $\lambda , \pi$, and $h$. 
The left eigenvector $\pi$ holds the stationary 
composition of the population, in the sense that 
\be{eq:pi}
\lim_{t\to \infty}
\frac{Z(t)}{\| Z(t)\|_1} = \pi \quad \text{with probability one,
conditionally on survival},
\ee
where $\| Z(t) \|_1  := \sum_{j \in S} Z_j(t)$ is the total population
size. 
This is due to the famous  
Kesten-Stigum theorem, see \cite{KeSt66} for the discrete-time original,
and \cite[Thm.~2, p.~206]{AtNe72}  and 
\cite[Thm.~2.1]{GeBa03} for  continuous-time versions.
Furthermore, with $R=(R_i)_{i \in S}$,
\be{eq:growthrate}
  \sp{\pi,R} = \lambda 
  = \lim_{t \to \infty} \frac{1}{t} \log  \| Z(t) \|_1
\ee
is the  asymptotic growth rate (or equilibrium mean fitness) of
the population. Here the first equality follows from the identity
$\lambda=\sp{\pi \A,\1}=\sp{\pi, \A\1}=\sp{\pi,R}$, and the second
is from \cite{GeBa03} and holds with probability one
in the case of survival.
Finally, the $i$-th coordinate $h_i$ of the
right eigenvector $h$ measures  the asymptotic 
mean offspring size of an $i$-individual, relative to the total size of the
population:
\be{eq:h}
h_i = \lim_{t\to \infty} \EE^i \big ( \|Z(t)\|_1 \big ) e^{-\lambda t}.
\ee 
For more details concerning  this quantity, see 
\cite{HRWB02} and \cite{GeBa03} (for the deterministic and stochastic
pictures, respectively).

\begin{figure}[h]
  \psfrag{t}{$t$}
  \psfrag{ttau}{$t-\tau$}
  \begin{center}
  \includegraphics{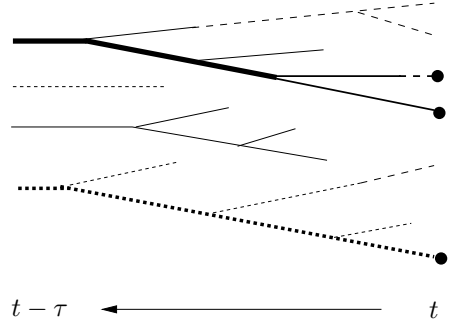}
  \end{center}
% exported with 60 percent
  \caption{\label{fig:backward} {\small The backward point of view. The
  various types are indicated by different line styles (solid, dashed and
  dotted).
  The fat lines mark the lines
  of descent defined by three individuals (bullets) picked from
  the branching population at time $t$. After coalescence of
  two such lines, the common ancestor receives twice the `weight', as
  indicated by the extra fat line; this motivates the factor
  $h_i$ in the ancestral distribution.}}
\end{figure}

In the above, we have adopted the traditional view on branching processes,
which is forward in time. It is less customary, but equally rewarding,
to look at branching populations backward in time. To this end,
consider picking individuals randomly (with equal weight) from the
current population and tracing their lines of descent backward in time
(see Fig.~\ref{fig:backward}). 
If we pick an individual at time $t$ and ask for the probability that
the type of its ancestor is $i$ at an earlier time $t-\tau$, the answer
will be $\alpha_i = \pi_i h_i$ in the limit when first $t \ti$ and then 
$\tau \ti$.
Thus the distribution $\alpha=(\alpha_i)_{i \in S}$ describes the
\emph{population average} of the ancestral types and is
termed the \emph{ancestral distribution}, see \cite[Thm.~3.1]{GeBa03}
for details. In contrast and for easy distinction, 
we will sometimes denote the stationary distribution of the forward
process, $\pi$, as the {\em present distribution}.

If we pick individuals from the population at a very late
time (so that its composition is given by the stationary vector $\pi$),
then the type process in the backward direction is the Markov chain with
generator $\G=(G_{ij})_{i,j \in S}$, 
$G_{ij} = \pi_j (A_{ji} - \lambda \delta_{ij}) \pi_i^{-1}$,
as first identified by Jagers \cite{Jage89,Jage92}. 
Unsurprisingly, its stationary distribution is the ancestral distribution, $\alpha$.

\section{An example: Sequence space model with single-peaked landscape.}
\label{sec:SPL}
\begin{figure}[h]
\begin{center}
 \includegraphics[width=.5\textwidth]{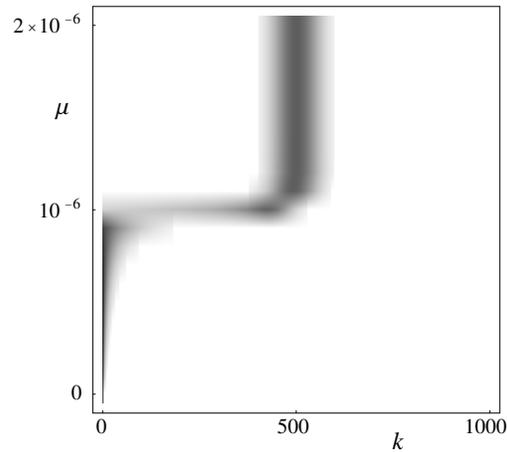}
\end{center}
  \caption{\label{fig:SPL} { \small The stationary distribution, $\pi$,
for the single-peaked landscape with  $L=1000$ and
$s=0.001$, as a function of the mutation rate per site, $\mu$.
Types are lumped into classes:  class $k$ contains
all sequences with $\#_1(i)=k$, where $\#_1(i)$ is the number
of 1's in sequence $i$. 
Grey levels correspond to $\sum_{i:\#_1(i)=k} \pi_k$ (darker shading
corresponds to larger values). The error threshold occurs at 
$\mu_0 \approx s/L=10^{-6}$.}}
\end{figure}

Assume now that $S= \{0,1\}^L$, i.e., the type of an individual is characterised by a  sequence of length $L$, of which every site may be either $0$ or $1$.
(This may be interpreted as a sequence of matches (0) or
mis\-matches (1) with respect to
a reference sequence 
of nucleotides or amino acids, for example.)
Assume that every site mutates at rate $\mu>0$ from $0$ to $1$ or vice versa,
independently of the other sites.
Assume further that $D_i = 1$ for all $i=(i_1,i_2, \ldots, i_L) \in S$, 
$B_{00\ldots 0} = 1+ s$, and $B_i = 1$
for $00\ldots 0 \neq i$; i.e., the reference sequence has a
selective advantage of $s>0$ over all others, 
which are equally unfit. 
This is a branching process version of what is known as the
{\em single-peaked landscape} of sequence evolution, which was 
introduced by Eigen \cite{Eige71} and has been discussed in numerous variants
ever since (see \cite{EMcCS89} and \cite{BaGa00} for reviews). 
Though not particularly realistic, it can serve as an instructive 
prototype model. 
Fig.~\ref{fig:SPL} shows its stationary type distribution, $\pi$,
as a function of the mutation rate. When there is little 
mutation, the population consists mainly of `fit' individuals; with
increasing $\mu$, more and more mutants come up, until the favourable
type is (nearly) lost, and the population turns into an 
(approximate) equidistribution
on $S$, thus losing its genetic structure. This transition occurs in a fairly
abrupt manner, around a value of $\mu \approx \mu_0=s/L$, and becomes
sharp (i.e., turns into a phase transition known as an {\em error threshold})
when $L \to \infty$. For a discussion of this and related threshold
phenomena, see \cite{HRWB02}.

\section{A two-type caricature.}
\label{sec:twotype}
\begin{figure}[h]
%export with 50 percent
  \centerline{\input{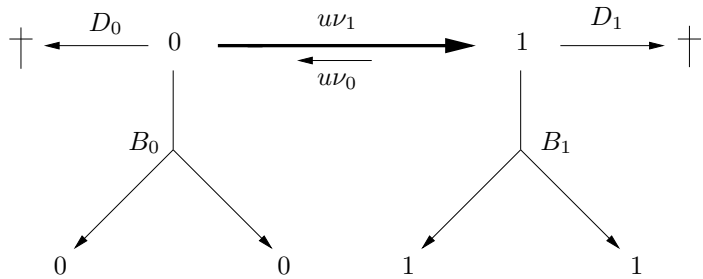}}
  \caption{\label{fig:twotype} {\small The two-type mutation-reproduction
   process.}}
\end{figure}

To capture the essence of the single-peaked model, let us condense it
radically. Assume that there are only two types, i.e., $S=\{0,1\}$,
with birth rates $B_0=1+s$, $B_1=1$, death rates $D_0=D_1=1$, and
mutation rates $U_{01}=u \nu_1$ and $U_{10}=u \nu_0$; cf.
Fig.~\ref{fig:twotype}. Here, $u>0$ scales the overall mutation rate, whereas
$\nu_0>0$ and $\nu_1>0$ ($\nu_0+\nu_1=1$) determine the asymmetry of mutation.
To mimic the single-peaked model, 
with the fit sequence corresponding to type 0 and all others 
collected into type 1, a simple choice is $\nu_0 = 1/L$ and $u=\mu L$;
it approximates the fact that, in the sequence picture, $\mu L$ is 
the total mutation rate 
away from $00\ldots 0$, and $\mu$ is the maximal rate at which an
unfit individual can mutate  back to the fit
type. For this choice, $\pi$ and $\alpha$ are shown in 
Fig.~\ref{fig:stat_backandforth} as a function
of $u$. Unsurprisingly, the stationary frequency of the fit type
decreases (almost linearly) with $u$, until it gets close to
zero (around $u_0=s$). In contrast, the ancestral
distribution consists almost exclusively of fit individuals until $u$
gets close to $u_0$,
and then declines to zero quickly.
(For $\nu_0 \to 0$, the curves turn into
piecewise linear ($\pi$) and piecewise constant ($\alpha$).) Obviously, for $u < u_0$, the present
distribution carries with it a `comet tail' of unfit mutants, which are present at any time, but
do not survive in the long run. They therefore rarely show up as ancestors,
nearly all of whom are fit.

\section{The Moran model with selection.}
\label{sec:moran}
Branching processes have very nice features mathematically, because individuals
reproduce  independently of each other. Biologically, however, this
independence can be quite unrealistic, because there is nothing to control population
size. At the other extreme, one considers models for populations of {\em fixed} size.
The
standard models in this context are the {\em Wright-Fisher model}  and
the {\em Moran model}. 
We will use the Moran model here because, in contrast to the Wright-Fisher
model, it has a continuous-time version, which leads to the coalescent
in a most immediate way; furthermore, it readily compares to our branching
process. Luckily, however, the choice is not essential,
since both the Wright-Fisher and the Moran models share the same
diffusion limit (up to a factor of 2, see Remark~2 below);
and it is only in this limit that the model can be tackled
explicitly anyway, as will become clear below.

Let us now embark on the counterpart of the  two-type branching model of the previous Section,
namely, the Moran model with two types, mutation and selection.

\begin{figure}[ht]
 \begin{center}
  \includegraphics[width=.6\textwidth]{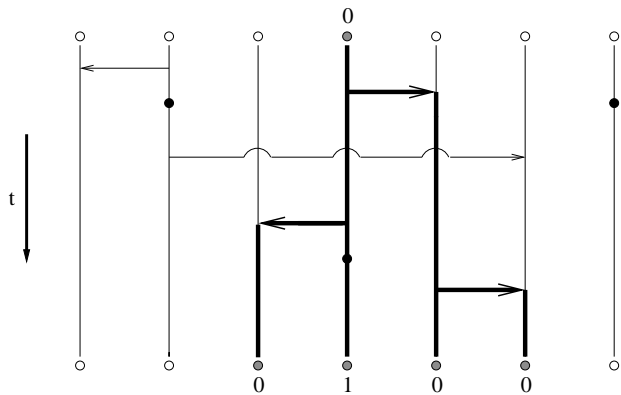}
 \end{center}
\caption{ \small The Moran model. The fat lines represent the descendants 
of the
single type-0 individual marked grey at the top ($t=0$). At the later time
(bottom), its descendants consist of four individuals, one of whom has
mutated to type 1 (mutation events are marked by blobs). 
Equivalently, the fat lines define the genealogy
of the four `grey' individuals at the bottom.}
\label{fig:moran}
\end{figure}

Moran models belong to the larger class of {\em interacting particle systems},
and are best explained in terms of their graphical representation, which, in our case,
goes as follows (see Fig.~\ref{fig:moran}). We have a fixed number of $N$ individuals, 
each characterised by a type $i \in S=\{0,1\}$, and represented by a vertical line; time runs from top to bottom. If an
individual gives birth, the offspring inherits the parent's type
and replaces an individual chosen
randomly from the population (maybe its own parent), thus keeping population size constant.
In the graphical representation, such events are displayed as arrows, with the parent
at the base and the offspring at the tip. Type-0 individuals reproduce at rate
$1+s$, type-1 individuals at rate $1$. 
This type dependence of reproduction rates leads to {\em selection}, which
means that, on average, unfit individuals are replaced by fit ones
more often than vice versa, thus leading to a net replacement of unfit
individuals by fit ones.
Independently of reproduction,
individuals may mutate, at rate $u \nu_1$ from type 0 to type 1, and at rate $u \nu_0$
in the reverse direction. Equivalently, we can say that every individual mutates at rate
$u$, and the ensuing type is $i$ with probability $\nu_i$, $i \in S$; 
one thus includes some `empty events' that, in effect, do not change 
the current type.
Mutation events are represented as blobs in the graphical representation.

The graphical representation is not meant to encode any spatial information --
all events are independent of the location of the individuals. 
Therefore, all relevant information is contained in the 
process $\{Y_0(t),Y_1(t)\}_{t \geq 0}$, where
$Y_i(t)$, $i \in S$, is the number of individuals of type $i$ at time $t$ 
(of course, $Y_0(t)+Y_1(t)=N$, but we retain both components for 
convenience here). If the current state is 
$(Y_0(t),Y_1(t))=(k_0,k_1)$ ($k_0=0,\ldots, N$),
the possible transitions are
\[
 (k_0,k_1) \to 
 \begin{cases}
 (k_0+1,k_1-1) \;  \text{at rate} \; u \nu_0 k_1 + (1+s) k_0 k_1/N  \\  
 (k_0-1,k_1+1) \; \text{at rate} \;  u \nu_1 k_0 + k_0 k_1/N
 \end{cases}
\]
(note that transitions to `impossible' states ($k_i \notin \{0,\ldots,N\}$)
occur at rate zero and are therefore excluded). 

The standard strategy to make the process tractable is 
to consider the normalised process 
$\{X_0(t),X_1(t)\}_{t \geq 0}:=\{Y_0(t),Y_1(t)\}_{t \geq 0}/N$,
and  take a diffusion limit, 
in which time is rescaled in units of $N$ generations, and the limit
$N \to \infty$ is taken so that $Ns \to \sigma$ and $N u \to \theta$, where $\theta >0$ and
$\sigma\geq 0$.
%\footnote{One should not wonder about the factor of 2:
%it is included for the sake of compatibility with the Wright-Fisher model, which shares 
%the same diffusion limit.}. 
In this limit, the stationary distribution of type frequencies, in the
forward direction of time, is well known: it is given by the density
\begin{equation}\label{eq:wright}
  f(x_0,x_1) = \frac{1}{C} x_0^{\theta \nu_0-1} x_1^{\theta \nu_1 - 1} e^{\sigma x_0},
 \end{equation}
 where $C$ is the normalising constant. Eq.~\eqref{eq:wright} is a classical result known
 as Wright's formula; see \cite[Ch.~4,5]{Ewen04} for a comprehensive review of diffusion theory in the context of population genetics.
 
 In the diffusion limit, the above particle representation is
no longer well-defined; a more
 advanced construction, known as the {\em look-down process} 
(\cite{DoKu96,DoKu99}, see also \cite[Ch.~5.2 and 5.3]{Ethe00})
 is required. In this tutorial introduction, we will only use the 
particle system
for the finite population, and derive from it a good motivation
for the backward construction in the diffusion limit.

\remar{Remark\ {1.}\ }{
In most of the literature, the Moran model with selection is defined
in a slightly different version,
in which mutation events can only occur on the occasion of reproduction
(see, e.g., \cite[Chap.~3.1]{Durr02}). But this {\em coupled} version
has the same diffusion limit as our parallel model, provided the
parameters are interpreted in a suitable way; actually, one
effect of the diffusion limit is to decouple mutation and selection,
which are both rare events in the scaling chosen. We therefore
prefer to start with the decoupled version right away.
}
 
 \remar{Remark\ {2.}\ }{
 In the literature,  a factor of 2 is often included in the mutation rates, selection
 intensities, and the time scale. The purpose of this is to make the Moran model comparable with
 the Wright-Fisher model, so that they share the same diffusion limit. More precisely,
 the factor of 2 comes from the fact that reproduction is tied to
 {\em ordered} pairs of individuals in the Moran model, but to unordered pairs in the
 Wright-Fisher model, cf.~\cite[p.~23]{Durr02}.
  Since we do not consider the Wright-Fisher model in this article,
 we will not stick to this convention, thus making our life easier.  Let us note for
 completeness that a second factor of 2 may appear if 
 populations are diploid (unlike the haploid populations considered here,
they carry two copies of the genetic information per cell,
which separate during reproduction, thus effectively duplicating the `population' size). 
 }
 
 \section{Neutral genealogies.}
\label{sec:neutral}
 The case without selection ($s=\sigma=0$),
  known as the {\em neutral} case in population genetics, is particularly simple.
  This is because reproduction rates do not depend on the type, and the
  reproduction and mutation processes may therefore be
superimposed independently. This allows one to go beyond the 
`frequency picture' considered in the previous Section,
 and also construct {\em genealogies}, i.e., answer the question:
 If I take a sample of $m$ individuals from a stationary population,
 which is the law that governs their joint history?

We will define a {\em genealogy} to consist of the
{\em genealogical tree}  and the {\em types} along the branches;
the genealogical tree, in turn, is given by the {\em tree topology} together
with the {\em branch lengths} (in units of time).
 If a realisation of the Moran model (in forward time)
 is given, the genealogy can be read off
 immediately by starting from the sampled individuals and tracing their 
ancestry 
backward (see Fig.~\ref{fig:moran}):
 Every time the tip of an arrow is encountered, that arrow is followed
backwards, and if it joins two lines in the genealogy,
 these lines merge. They thus give rise to a {\em coalescence event}, 
that is,
 two individuals go back to a common ancestor. The crucial point now is that
 realisations of genealogies can be constructed without constructing realisations of the forward
 process first, by a two-step procedure:
 
 Start from our $m$ individuals  and construct their
 genealogical tree, going backward in time and {\em ignoring the types}.
 To this end,
 recall that,  in the graphical representation of
 the forward process, arrows appear at rate $1/N$  between
 any ordered pair of individuals (on the original time scale).
 Hence, arrows that  lead  to coalescence events  appear
 at rate $1/N$ per ordered pair of lines in the genealogical tree. 
The number of lines currently in the
 tree is therefore a  death process, which starts at $m$, decreases
 in unit steps at rate $n(n-1)/N$ if the current state is $n$, and ends
in the absorbing state 1; at every death event, a random pair of lines
is chosen to merge. The individual at 
the root of the tree is the
{\em most recent common ancestor (MRCA)} of the sample.
This verbal description
defines the law, and an easy method of 
sampling,
 for  genealogical trees. Rescaling of time turns the death
rates into $n(n-1)$; the resulting stochastic process is known
as the {\em coalescent process}, and goes back to Kingman 
\cite{King82a,King82b}.
The types along the branches may then be determined in a second step,
 by picking the type of the most recent common ancestor from the stationary
 distribution \eqref{eq:wright}
and running the mutation process forward in time along the genealogical tree 
(where, of course, both descendants inherit the parent's type at a
coalescence event).

\section{The ancestral selection graph.}
\label{sec:ASG}
As mentioned above, the simplicity of the above construction 
relies on the fact that,
in the neutral case, the  mutation process may be
superimposed independently onto the
reproduction process, and hence the genealogical tree. 
But this independence breaks down as soon as there is  selection.
For this reason, the
coalescent with selection has been considered intractable for 
more than a decade.
Only within the last decade, Neuhauser and Krone \cite{KrNe97,NeKr97}
and Donnelly and Kurtz \cite{DoKu99} discovered constructions  
that overcome this problem. We will stick to the Neuhauser-Krone
framework here, but 
the fundamental idea, in both constructions, is to  again establish  some kind of separation of
the genealogical and the mutation processes. This is achieved by the
following trick.
Start with the forward particle system, and ignore the types.
Decompose the original (type-dependent) reproduction process (before
rescaling of time)
into two independent processes,
represented by two types of arrows: `neutral' arrows that appear at rate 1 on every line
regardless of the type (as in the neutral case, hence the name); 
and `selective' arrows that appear at rate $s$, again {\em on every line regardless of the type},
but with the additional convention that the latter arrows are only  `used'  
if they emanate from a fit individual; otherwise, they are ignored. 
In contrast, neutral
arrows are used in either case. Put differently, neutral arrows
represent {\em definitive birth events} for everyone, whereas selective arrows
correspond to  {\em potential birth events}, to be used by  fit individuals
only. If
we now reintroduce the types (by defining the initial types and adding in 
mutation events),
the resulting process is equivalent to the original 
Moran model with selection (forward in time). Backward in time, it
allows the construction
of the genealogy of a sample taken from the stationary distribution, but now
three steps are required, rather than two as in the neutral case.
To explain the idea, let us give
a verbal description, still starting from a given realisation of the collection
of arrows in a finite particle system, but without knowledge of types; later, we will free ourselves from concrete realisations, and pass to the diffusion
limit.

\begin{enumerate}
\item[(A1)]
In a first step (see Fig.~\ref{fig:asg}, left panel), construct the so-called 
{\em ancestral selection graph (ASG), without types.} To
this end, start from
a sample of $m$ individuals from the present population,
 {\em with the types unknown}, 
and {\em run time backward}. This is done as follows.
Given the realisation of the pattern of arrows, trace back 
the lines; every time you meet a
neutral arrow,  proceed in the familiar way, thus producing a
coalescence event.
However, if a selective arrow is encountered, there is a potential
birth event which cannot be resolved yet since the types are still unknown.
Depending on whether the arrow has or has not been used,
there are two possible
parents: the line at the base of the arrow (the {\em incoming branch}), or
the line at the tip  (the {\em continuing branch}). 
To keep track of both possibilities, the incoming branch is attached to
the graph,
which results in a {\em branching event}. (We ignore the
possibility that the incoming branch may  already be contained in
the graph, since the probability for this vanishes 
in the diffusion limit.)
The resulting {\em branching-coalescing graph} is known as the
ASG (without types);   all 
{\em possible genealogical trees}
are {\em embedded} in it. 
When the number of lines reaches 1 for the
first time, we have found the so-called
{\em ultimate ancestor (UA)} of the population (not to be
confused with the MRCA, see below).  Note, however, that,
going back further in time, the  line emanating from the UA
will, sooner or later,  branch out again; the
process may be continued indefinitely.

\begin{figure}[ht]
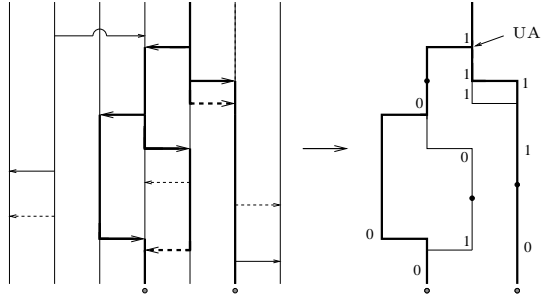

%export with 30 percent
        \centerline{\input bild111.pstex_t}
\caption{{\small Constructing the ASG. Left: Particle system with 
neutral arrows
(solid) and selective arrows (dashed). Embedded (as bold lines)
is the ASG without types, as obtained when going backward from the two
`grey' individuals sampled from the present population. Right:
 Starting from
the UA, the mutation process is superimposed on the ASG, which may then
be resolved into real (bold) and virtual (thin) branches
according to Fig.~\ref{fig:resolve}. The collection of bold lines
(along with their types) constitutes the genealogy of the 
grey individuals.
}}
\label{fig:asg}
\end{figure}

\item[(A2)]
In the second step (see Fig.~\ref{fig:asg}, right panel), 
the type of the ultimate ancestor is drawn
from the stationary distribution of the forward process. 
(It is not immediately obvious
that the UA's type should  follow this distribution; the reason it
does lies in the fact that
the time when the ASG reaches size 1 is independent of the
type, see the arguments in \cite{StDo03} and \cite[Thm.~8.2]{DoKu99}.
Note that, even for finite
$N$, the stationary distribution is available in closed form
\cite{WBS95}; but it simplifies considerably in the
diffusion limit.) 
Starting from the UA, the mutation process is then superimposed on
the graph, running {\em forward in time}. As usual,
both descendants inherit the
parent's type at a  coalescence event. 
And now that the types are known, the branching events 
 can  be resolved as one goes along (see Fig.~\ref{fig:resolve}): 
If the incoming 
branch is fit,  the selective birth event has indeed taken
place, and the incoming branch is {\em parental} to
the descendant individual (the
one below the tip of the arrow); 
the continuing branch is 
called {\em virtual}. If the incoming branch
is, however, unfit, then
the birth event in question was fictitious, the 
incoming branch is  virtual, and the continuing branch is the parental
one. Note that, as a result, the descendant individual  is unfit
if and only if both the incoming and the continuing branches are unfit.
The result of step (A2) is known as the {\em ASG with types}.

\begin{figure}[ht]
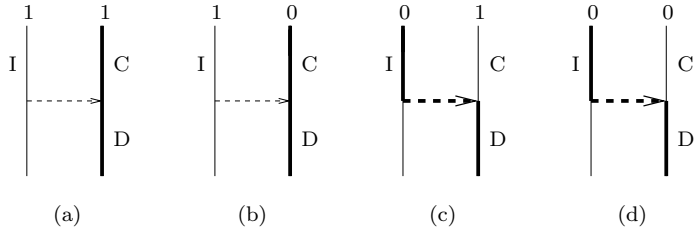

%export with 50 percent
        \centerline{\input bild15.pstex_t}
\caption{{\small Resolution of potential birth events. I incoming branch,
C continuing branch, D descendant. The descendant line and its
parental branch are marked bold.}}
\label{fig:resolve}
\end{figure}

\pagebreak[4]

\item[(A3)]
In the third and final step (see again  Fig.~\ref{fig:asg}, right panel), 
the genealogy is extracted by 
moving backward in time once more. Starting at the sampled individuals,
we trace their ancestry backwards along their parental lines 
identified in the previous step. The collection of all lines that are
{\em parental to the sampled individuals} constitute the genealogical tree;
they are also called {\em real branches}. All other lines are
virtual (in particular,
lines that are parental to virtual branches are themselves
virtual), and are now removed from the picture to yield the genealogy. 
Note  that  the 
MRCA of the genealogical tree need not
coincide with the UA of the ASG; it may be younger.
\end{enumerate}
The diffusion limit simplifies this construction in two
respects. In (A1), the probability for the incoming branch to
be one that is already within the ASG
vanishes in the limit. And in (A2), it provides us
with the simple explicit form of the stationary distribution \eqref{eq:wright}
from which
to pick the UA's type. Forgetting about concrete realisations and considering
that, after rescaling of time, selective arrows land on every line at
rate $\sigma$, we can construct the {\em ASG in the diffusive limit}
as the following analogue of (A1)--(A3):
\begin{enumerate}
\item[(A1')] 
Construct the number of lines in the graph as a birth-death process,
starting at $m$, with death 
rate $n(n-1)$ and birth  rate $n\sigma$ when there are 
currently $n$ lines. From this, construct the ASG without types by
choosing a random pair to coalesce
at every death event, and choosing a random line to branch into two
at every birth event.
Stop when there is only one line left (this will almost surely
happen in finite time \cite[Thm.~3.2]{KrNe97}).
\item[(A2')]
Pick the type of the ultimate ancestor from
the stationary distribution \eqref{eq:wright}. Run the mutation process
down the graph (at rates $u \nu_0$ and $u \nu_1$) and resolve branching
events as you go along, according to Fig.~\ref{fig:resolve}.
\item[(A3')]
Extract the genealogy, as in (A3).
\end{enumerate}
An explicit algorithm is given in \cite[Alg.~3.1]{StDo03}.

\section{Tracing back single individuals.}
\label{sec:single}
After describing the general idea of the ASG, we now turn to the simpler 
situation that arises if we trace back the ancestry of
{\em single} individuals, picked
from the stationary (present) distribution, rather
than the genealogies of {\em samples} of several individuals.
For this purpose, a
simplified construction is available, which was recently presented by
Fearnhead \cite{Fear02}. For a complementary approach, which relies on
diffusion theory alone (without reference to genealogies), the
reader is referred to
Taylor \cite{Tayl06}; we will restrict ourselves to the genealogical approach
of \cite{Fear02} here.
From now on, we will directly work with the diffusion
limit.

Even if only a single individual is sampled, its ancestral line will
branch out at some stage; we will therefore still require the law for
{\em samples}. Consider, therefore, a sample of size $m$ (which,
for the moment, comes from the present population).
One may either describe it as an {\em ordered sample} 
$\c =(c_1, c_2, \ldots, c_m)$, 
 where $c_{\ell}$ is the type of the
$\ell$-th individual ($1 \leq \ell \leq m$),
or as an {\em unordered sample} $\m=(m_0,m_1)$,
where $m_i$, $i \in S$, is the number of individuals of type $i$
(of course, $m_0+m_1=m$).
Since there is no spatial information in our model, the law of our
sample is invariant under permutations of the  individuals. More
precisely, if $p^*(\c)$ is the probability of getting the ordered
sample $\c$ when sampling $m$ individuals, we have 
$p^*(\hat \c)=p^*(\c)$ if $\#_i(\hat \c) = \#_i(\c)$, $i\in S$,
where $\#_i(\c)$ is the number of individuals of type $i$ in 
$\c$; this
property is known as {\em exchangeability} in probability theory.
One could therefore work with unordered samples; for
our purpose, however,
it is convenient (and compatible with \cite{Fear02})
to work with ordered samples, but to retain a slightly abusive
notation reminiscent of unordered samples. Namely, we will denote
by $p(\m)$ the probability of {\em any} ordered sample that has
$m_0$ 0-individuals and $m_1$ 1-individuals.
More precisely, $p(\m)$ satisfies $p^*(\c) = p(\m)$ 
for every $\c$ with $\#_0(\c)=m_0$
and $\#_1(\c)=m_1$.  By
Wright's formula \eqref{eq:wright}, we have $p(\m)=\EE(X_0^{m_0} X_1^{m_1}) = 
\int_{\X} x_0^{m_0} x_1^{m_1} f(x_0,x_1) dx_0 dx_1$, where
$\X := \{(x_0,x_1) \mid x_0,x_1 \geq 0 \; \text{for} \; i \in S, x_0+x_1=1\}$.
For later use, let us also define
\begin{equation}\label{eq:add_sample}
  p(j \mid \m) := \frac{p(\m+\e_j)}{p(\m)}, \quad j \in S,
\end{equation}
where
$\e_0:= (1,0)$ and
$\e_1 := (0,1)$; in words, $ p(j \mid \m)$ is  the probability of obtaining
an individual of type $j$ if we have already drawn a
sample $\m$ and pick one additional
individual from the 
%\eb{(same realization of the?} 
stationary distribution. 

Armed this way, let us now describe the construction of the ancestral
line of a single individual. A further
simplification will result from the fact that the types
will be included throughout.
Modifying the construction (A1')--(A3'), one proceeds in the
following way (see \cite{Fear02} and \cite{StDo03} for  details):

\begin{enumerate}
\item[(B1)]
Pick an individual (i.e., a sample of size one) from the present
population. Choose its type according to Wright's formula \eqref{eq:wright},
i.e., $p(\e_i) = \EE(X_i)$,  $i \in S$.

\item[(B2)]
Given this type, construct the ASG {\em including the types}.
This only requires a single step because, given the initial type,
the mutation process may be run backward, and,
when a branching event is encountered, the type of the descendant is
already known --- and hence the type of
the parental branch (be it `incoming' or `continuing'), 
because it coincides with  the type
of the descendant. This way,
potential birth events can 
be resolved on the way back already, thus contracting steps (A1') and (A2').
In particular, this applies to the real branch
(there is exactly one real branch throughout --
the line parental to the single individual picked in step (B1)).
What needs to be assigned at a branching event is
the type of the virtual branch that appears;
this can be done by using Fig.~\ref{fig:asgtrans},
and will be formalised below. 
Let us only conclude here that the ASG boils
down to a process with states $(i;\n) \in  S \times \ZZ_{\geq 0}^2$,
where $i$ denotes the type of the real branch, and $\n=(n_0,n_1)$
collects the numbers of virtual branches of type $0$ and $1$,
respectively.
\end{enumerate}

Ultimately, to determine the ancestral distribution of types analogous to
that of the branching process, we will only need the real branch -- but
to determine its type, the virtual ones are indispensable, as will become
clear in a moment. 

%Our aim will be to determine the stationary distribution $\pi$
%of this process. Marginalising over the virtual branches will then
%give the stationary distribution of the ancestral type;
%i.e., $\sum_n \pi(i;n)$ is the stationary probability that the
%ancestral line has type $i$. This is the quantity that corresponds to
%$\alpha_i$ in the branching process.

\section{Transitions and rates of the ASG, including the types.}
\label{sec:rates}
Let us now make precise
step (B2) by recapitulating from
\cite{Fear02} the definition of the ASG in terms of its 
transitions and the corresponding rates; for a detailed
derivation, we refer the reader to \cite{Fear02} and
\cite[Alg.~3.2]{StDo03}. Let us only recall here that
working backward in time
(relative to the stationary measure $p$ of ordered samples) 
is analogous to
constructing  the generator for the time-reversed version of
a continuous-time Markov chain on a state space $E$, as given
by $\bar Q = (\bar Q_{ij})_{i,j \in E}$, where 
$\bar Q_{ij} = \pi_i^{-1} Q_{ji} \pi_j$  if $Q=(Q_{ij})_{i,j \in E}$
is the generator of the forward process and $\pi=(\pi_i)_{i \in E}$
is the corresponding stationary distribution; cf.~\cite[Ch.~3.7]{Norr97}
for some introductory material on time reversal;
and compare the related  time-reversed
branching process in Sec.~\ref{sec:branching}.
Let us also note that, for ease of exposition, we will include `empty events' 
in the context of mutation (as in Sec.~\ref{sec:moran}).

\begin{figure}
%export 45 percent
\centerline{\input{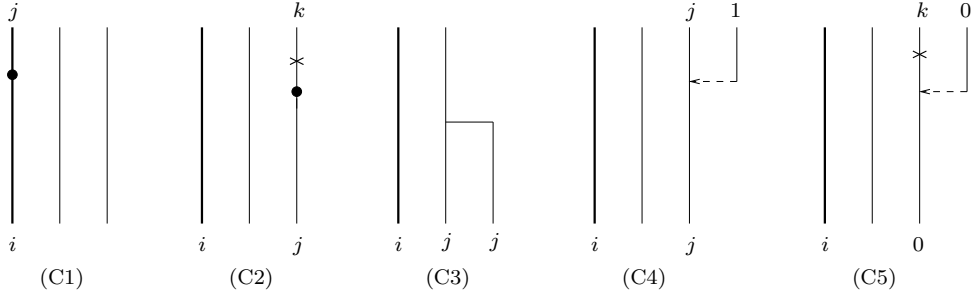}}
\caption{\label{fig:asgtrans} {\small 
Transitions of the  ASG, including the types.
Transitions (C1)--(C5) out of $(i;\n)$ in the ASG with types.
The bold line is the real branch, the thin lines represent virtual branches.
Blobs mark mutation events, crosses mark 
branches whose types are obtained as a random draw according to
\eqref{eq:add_sample}, and may be removed. In (C4), the continuing 
branch is parental; 
in (C5), the incoming branch is parental. Backward time runs from
bottom to top.}}
\end{figure}

Given that the
current state is $(i;\n)$ and going backward in time,
five  types of events can occur  (cf.~Fig.~\ref{fig:asgtrans});
note that $(i;\n)$ corresponds to a sample $\n+\e_i$. 

\begin{enumerate}
\item[(C1)] {\em Mutation of the real branch:} $(i;\n) \to (j;\n)$, $j \in S$,
occurs at rate $\theta \nu_i p(\n+\e_j)/p(\n+\e_i)$, in accordance with the
above reversion recipe.

\item[(C2)] {\em Mutation of a virtual branch}:  
$(i;\n) \to (i; \n+\e_k-\e_j)$, $j,k \in S$, occurs at rate 
$n_j \theta \nu_j p(\n+\e_i-\e_j+\e_k)/p(\n+\e_i)$. 
The factor $n_j$ comes from the fact that, in the corresponding
forward transition, $(i,\n)$ is the target state and contains
$n_j$ virtual branches
of type $j$, each of which may have arisen by mutation 
(recall that $p(\m)$ is the probability of {\em any ordered} sample
of composition $\m$). 
Let us note for later use
that, by \eqref{eq:add_sample}, the rate may be rewritten as 
$(n_j \theta \nu_j p(\n+\e_i-\e_j) /p(\n+\e_i)) \times$ $p(k \mid \n+\e_i-\e_j)$,
that is, the type of the mutated branch can be obtained
as a random draw from the stationary population, given the
other branches in the sample.

\item[(C3)]
 {\em Coalescence of two branches of type} $j$, $j \in S$:
$(i;\n) \to (i; \n-\e_j)$ occurs at rate $(n_j + \delta_{ij}) (n_j + \delta_{ij}-1)
p(\n+\e_i-\e_j)/p(\n+\e_i)$. 
The combinatorial factor reflects
the fact that coalescence events may involve both real and virtual branches, 
but can
only occur between like types; for type $j$, there are
$(n_j + \delta_{ij}) (n_j + \delta_{ij}-1)$ ordered pairs of type $j$.

\item[(C4)]
{\em Branching to an unfit incoming branch} (see also Fig.~\ref{fig:resolve}, (a)
and (b)): 
$(i;\n) \to (i; \n+\e_1)$ occurs at rate
$(n_j + \delta_{ij}) \sigma p(\n+\e_i+\e_1)/p(\n+\e_i)$, for every $j \in S$.
Here, the continuing branch (regardless of its type) 
is parental, and its type is thus the same as before the branching event.

\item[(C5)] {\em Branching to a fit incoming branch}
(see also Fig.~\ref{fig:resolve}, (c) and (d)): 
$(i;\n) \to (i; \n+\e_k)$, $k \in S$, occurs at rate ($n_0+\delta_{i0}) \sigma 
p(\n+\e_i+\e_k)/p(\n+\e_i) = (n_0+\delta_{i0}) \sigma p(k \mid \n+\e_i)$.
Here, the incoming branch is parental; the type of the
continuing branch may be obtained as a random draw conditional on the
remaining sample, similar to the situation in (C2).

\end{enumerate}

The form of the mutation rate in (C1) is essential to understanding why 
we must keep track of the virtual
branches at all, if we are, ultimately, only
interested in the type of the real branch: the mutation process on the
real branch depends on the full state of the process, including the virtual
branches. Considering the real branch in isolation would, therefore,
not lead to a {\em Markov} process. Indeed, the achievement of the
ASG construction, and further simplifications like the {\em common
ancestor process}
below, lies just in the  construction of a process with a state
space
that is `as small as possible', with as few transitions as possible, 
while still
retaining the Markov property. Indeed, the lines in the ASG  
only represent a vanishing proportion of the population (which is
infinite in the diffusion limit!); loosely speaking,
they may be considered a `representative sample'.

\section{The common ancestor process.}
\label{sec:pruned}
Since we are only interested in the history of one individual,
genealogical trees do not matter. This allows for a further simplification.
According to \cite[Thm.~1]{Fear02},  virtual branches whose types
have the law of a random draw  from
the stationary population (conditionally on the types of the
remaining sample, as in events (C2) and (C5))
may be removed; the remaining ones still perform a Markov
process defined by the events and rates (C1)--(C5). The intuition behind this
is the following. Let $k$ be the type of a virtual branch in question,
in a sample $(\m + \e_k)$. If $k$ is distributed according to
$p(k \mid \m)$, then it contains no information about the history
of the remaining sample, as suggested by the following thought experiment.
Consider you draw  $m+1$ individuals from
the population, but determine only the type of $m$ of them, resulting
in the (known) sample $\m$.
Then $p(k \mid \m)$ is the law of the type of the unknown individual. 
Clearly, then, 
the history of the known individuals
is  an ASG, regardless of the unknown type.

Consider now events (C2) and (C5). Since, in both cases, the type of a virtual
branch has a distribution of the form $p(k \mid \m)$, it can safely
be removed from the ASG without influencing the other branches (and,
in particular, the history of the ancestor). 
Removal of the mutated virtual branch  in
(C2) turns this transition  into one that is indistinguishable from a
coalescence event of two virtual branches, (C3). Step (C5) may be removed 
altogether, since removal of
the newly-created  virtual branch turns the transition into
an `empty' one. As a consequence, virtual branches can now only arise
due to (C4) and are all unfit. Therefore,  $n_0=0$ throughout,
and we set $n:= n_1$ in what follows.
We thus arrive at what was termed the
{\em common ancestor process (CAP)} in \cite{Fear02}, and
defined by its transitions out of $(i;\n)$:

\begin{enumerate}
\item[(P1)] {\em Mutation of the real branch}: $(i;\n) \to (j;\n)$, $j \in S$,
at rate $\theta \nu_i p(\n+\e_j)/p(\n+\e_i)$;

\item[(P2)] {\em Removal of a virtual branch (by coalescence or mutation)}:
$(i;\n) \to (i; \n-\e_1)$ at rate $[(n_1 + \delta_{i1})(n_1+\delta_{i1}-1)
+ \theta \nu_1 n_1] p(\n+\e_i-\e_1)/p(\n+\e_i)$;

\item[(P3)] {\em Branching}: $(i;\n) \to (i; \n+\e_1)$ at rate 
$\sigma(n_1+1) p(\n+\e_i+\e_j)/p(\n+\e_i)$.
(Note the typo in this rate in 
\cite{Fear02}: the factor $(n_1+1)$
is missing.)
\end{enumerate}

\begin{figure}[ht]
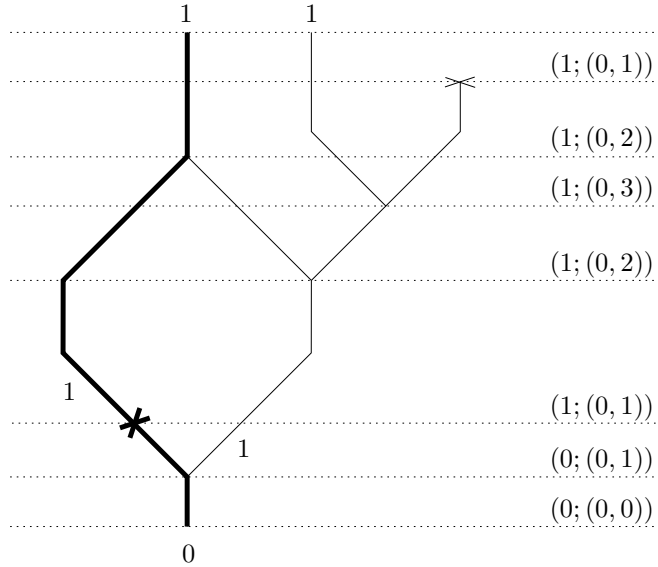

        \centerline{\input prune1.pstex_t}
\caption{{\small The common ancestor process. The solid line is the real branch,  thin lines
are virtual branches, numbers on lines mark their type.
Backward time runs from bottom to top. Mutation
events are marked by crosses; they lead to a change of type on the real
branch, but to extinction on virtual branches. Virtual branches are all
unfit. The state of the process between events is given to the right.}}
\label{fig:pruned_alp}
\end{figure}

The process starts in state $(i;(0,0))$ with probability $p(\e_i)$.
A realisation is shown in Fig.~\ref{fig:pruned_alp}.
Note   that the above notation of the states is somewhat
redundant (the support of the process is 
$E=S \times \{0\} \times \ZZ_{\geq 0}$), but we retain it here for 
compatibility with the `full' ASG.

\section{The stationary distribution of the common ancestor process.}
\label{sec:stationary}
By
Thm.~3 and Lemma 1 of \cite{Fear02}, the stationary distribution $a$ of the
CAP  is given by
\begin{equation}\label{eq:a}
  a(i;\n) = A_n(i) p(\n+\e_i), \quad (i;\n) \in E,
\end{equation}
where $A_n(0) := \prod_{j=1}^n \lambda_j$, $A_n(1) := A_n(0) (1-\lambda_{n+1})$,
and $ \lambda_j := \lim_{k \to \infty} \lambda_j^{(k)}$ ($j \geq 1$),
where the $\lambda_j^{(k)}$ ($k \geq 0$) are defined by $\lambda_{k+1}^{(k)}=0$, 
together with the recursion
\begin{equation}\label{eq:lambda}
   \lambda_{j-1}^{(k)} = 
   \frac{\sigma}{j+\theta + \sigma - (j+\theta \nu_1) \lambda_j^{(k)}}\,.
\end{equation}
The proof is by direct, though slightly cumbersome, verification;
an alternate proof, which uses advanced diffusion theory rather
than the ASG, is given in \cite{Tayl06}. 
But a nice interpretation  is also available \cite{Fear02},
which turns \eqref{eq:a} into the following rule for
simulating from the distribution $a$  \cite{Fear02}.
 Choose an individual at random from
the present
distribution \eqref{eq:wright}. If the individual is fit,
call it the real branch (i.e., the ancestor). If the individual 
is unfit, then with probability
$1-\lambda_1$ call it the ancestor; otherwise,  call it a virtual
branch, and  pick another individual from the present distribution.
Repeat this (i.e., if the $j$'th
individual chosen is unfit, then with probability $1-\lambda_j$
it is the  ancestor, otherwise  draw another one)
until you find the ancestor. When the
ancestor has been found, the state $(i;\n)$ is given by the type $i$ of the
ancestor, and the number $n_1$ of virtual branches (and $n_0=0$).

Marginalising over the virtual branches finally results in
the stationary distribution of the ancestral type; more precisely,
$a_i := \sum_{n\geq 0} a (i;n)$ is the stationary 
probability that the
ancestor has type $i$. Clearly, this is the desired analogue of
$\alpha_i$ in the branching process.

\section{An example.}
\label{sec:example}
\begin{figure}[h]
  \begin{center}
  \includegraphics[width=.45\textwidth]{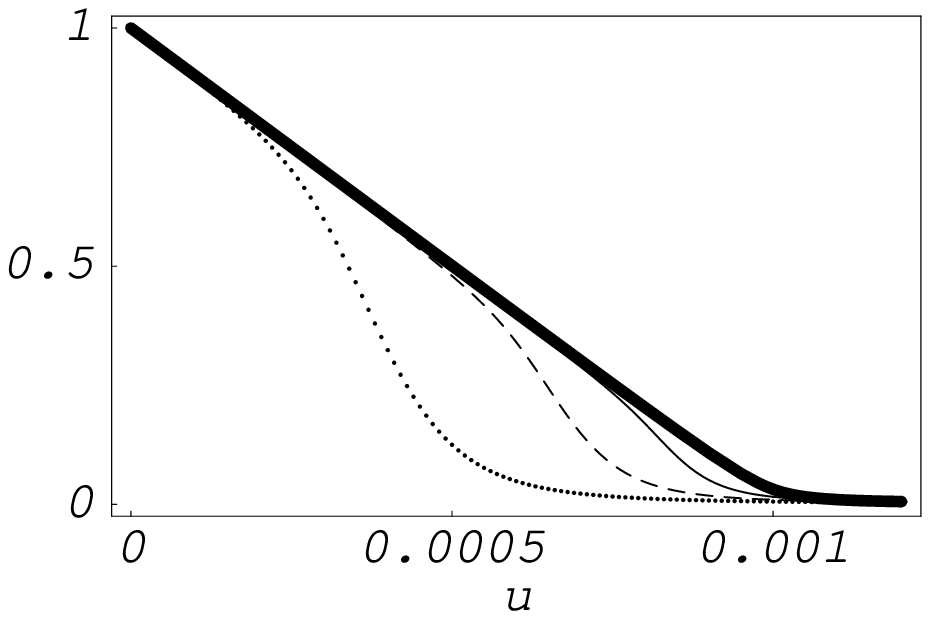}
  \quad 
  \includegraphics[width=.45\textwidth]{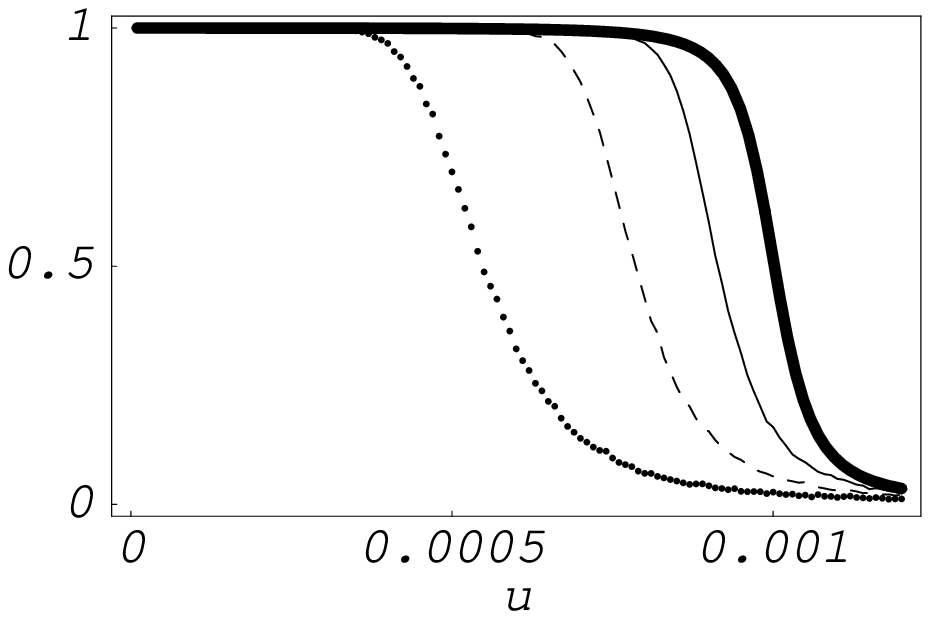}
  \end{center}
  \caption{\label{fig:stat_backandforth} 
{\small The present and ancestral  
frequencies of the fit type in the two-type branching process and the
corresponding Moran model with selection, as a function of
the mutation rate. Parameters:
$s=0.001$, $\nu_1=0.999$. Left: Expected frequency of fit individuals in the stationary
{\em present} distribution.
Bold line: $\pi_0$, of the branching
process; thin lines: $p(\e_0)$ in the Moran model, for $N=10000$ (dotted),
$N=30000$ (dashed), and $N=100000$ (solid line),
according to Wright's formula \eqref{eq:wright}.
 Right: Expected frequency of fit 
individuals in the ancestral distribution. Bold line: $\alpha_0$, of the branching
process; thin lines: $a_0$ in the Moran model, for $N=10000$ (dotted),
$N=30000$ (dashed), and $N=100000$ (solid line). Here, $a_0$
is approximated by simulating
according to the recipe resulting from \eqref{eq:a},
and averaging over 10000 realisations.
The $\lambda_j$ required in the simulation are approximated by
setting $\lambda_{500}^{(499)}=0$, and using the recursion
\eqref{eq:lambda}.
%(As revealed by simulation studies, our range of parameters is well
%within the range of validity of the diffusion approximation, see,
%e.g., \cite{pfaff}; but the approximation is well-known to break
%down if $s$ is large.)}
} }
\end{figure}

Recall the two-type branching caricature of the single-peaked model
(Sec.~\ref{sec:twotype}), and let us compare it to its Moran
analogue. For the values of $s$ and $\nu_0$ used before,  
Fig.~\ref{fig:stat_backandforth} shows the stationary frequencies of
fit individuals in the present and ancestral population, as a function
of $u$, and for various population sizes (that is, we take the approximating
diffusion defined by the choice  $\sigma=Ns$, $\theta=Nu$,
for given $s$ and $u$). 
More precisely, the Figure depicts the expected frequency
of fit individuals in the present population, $p(\e_0)=\EE(X_0)$ 
as given by Wright's formula;
and the expected frequency of fit ancestors  of individuals 
sampled randomly from the present population, as given by
$a_0$. 
Comparing these with
the corresponding stationary frequencies $\pi_0$ and $\alpha_0$
in the branching
process suggests that the stationary distribution of the
branching process is
related  to another  $N \to \infty$ limit of the Moran model, which
differs from the diffusion limit.
Indeed, the Moran model (forward in time)
also has a law of large numbers: Consider a sequence of Moran models
{\em with fixed $s$ and $u$}, and increasing $N$, {\em without rescaling
of time}. Denote the
corresponding normalised processes by  
$\{X_0^{(N)}(t),X_1^{(N)}(t)\}_{t \geq 0}$,
where the upper index is now introduced to denote dependence on
population size. 
One can then show that, for $N \to \infty$,
the  sequence of generators of the  processes 
$\{X_0^{(N)}(t),X_1^{(N)}(t)\}_{t \geq 0}$
converges, in a suitable sense, to the generator of the 
deterministic process defined by the system of ordinary differential
equations
\begin{equation}\label{eq:LLN}
  \dot p_i(t) = (R_i - \sp{R,p}) p_i(t) 
  + \sum_{j: j \neq i} ( p_j(t) U_{ji}- p_i(t) U_{ij}),
\quad i \in S,
\end{equation}
where $p_i(t)$ is the relative frequency of type $i$ at time $t$
(details will be given elsewhere). By Thms.~1.6.1 and 4.9.10 of
\cite{EtKu86}, this implies weak convergence of the stationary
distributions of $\{X_0^{(N)}(t),X_1^{(N)}(t)\}_{t \geq 0}$
to the stationary distribution of \eqref{eq:LLN}. But the latter, in
turn, is easily seen to coincide with the stationary distribution $\pi$
of the branching process (see, e.g., \cite{BaGe06}).
(A corresponding relationship for the backward process, suggestive as it
may seem, remains to be formally established.)
%\eb{Gilt fuer einzelne
%Realisierung, also erst recht fuer den Erwartungswert.}

It is interesting to observe how the convergence of the
stationary distribution manifests itself 
(see Fig.~\ref{fig:stat_backandforth}). 
The Moran  curves follow the branching curves closely
for small $u$ and then break off; this happens at values of $u$
that approach $u_0=s$ with increasing population size.
Therefore, the branching result is an excellent  approximation even
for small and moderate-sized  populations, as long as mutation rates are small.

These observations complement previous studies 
\cite{NoSc89,WBS95}, which only considered
the forward direction of time (they actually originated well before the 
ancestral selection graph). Let us  note that the transition
(in both branching and Moran models) becomes much steeper when the
mutation rates are  even more asymmetric than assumed here;
the behaviour can then be
interpreted as an error threshold for finite populations \cite{WBS95},
but this is not our  concern here.

\section{The virtual branches.}
\label{sec:virtuals}
\begin{figure}[h]
  \begin{center}
  \includegraphics[width=.45\textwidth]{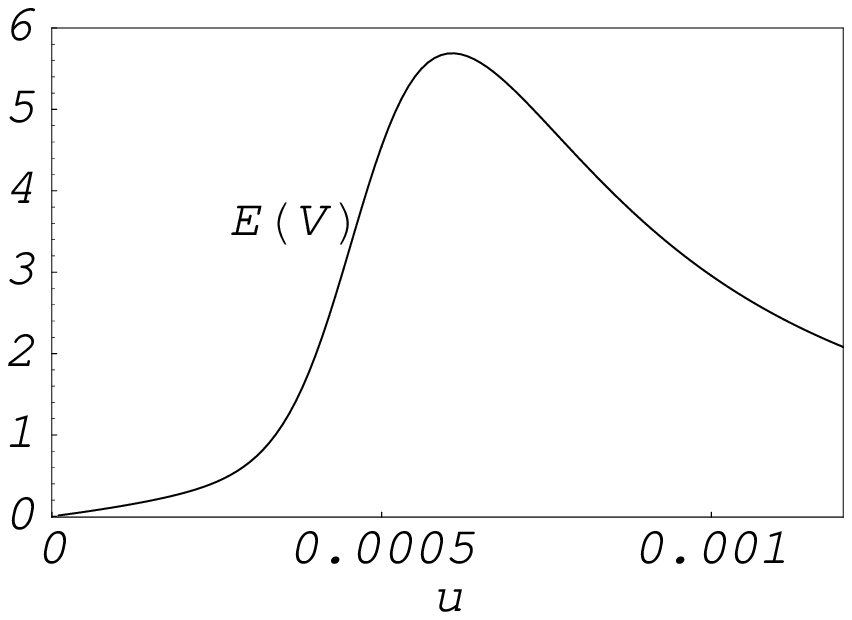}
  \quad
  \includegraphics[width=.45\textwidth]{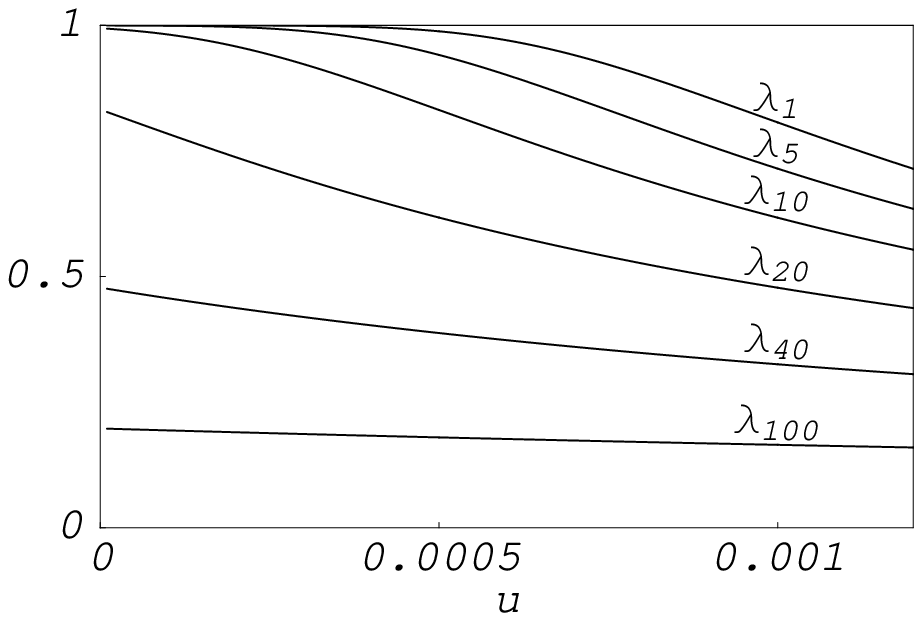}
  \end{center}
  \caption{\label{fig:virtuals} {\small 
 The virtual branches and how they
  originate. The situation corresponds to
          the dotted curve in Fig.~\ref{fig:stat_backandforth}, 
          i.e., $N=10000$,  $s=0.001$,
         $\nu_1=0.999$. 
  Left panel: Expected number of virtual
          branches, as a function of the mutation rate.
          The number of virtual
          branches, $V$, is simulated as in 
          Fig.~\ref{fig:stat_backandforth} (right), and averaged over
          10000  realisations to obtain an approximation
          of its expectation, $E(V)$.
 Right panel: 
 The $\lambda_j$ as a function of the mutation rate, 
calculated as in Fig.~\ref{fig:stat_backandforth} (right).}}
\end{figure}

The transition described above is intimately 
connected with the
number of virtual branches -- in fact, it is accompanied by an
explosion of the expected number of virtual branches, see
Fig.~\ref{fig:virtuals} (left panel). This can only be understood in the light
of how the $\lambda_j$ depend on $u$. As  shown in
Fig.~\ref{fig:virtuals} (right panel), $\lambda_1$ is very close to 1 at $u=0$,
and the $\lambda_j$ decrease with $u$, as well as $j$. 
The shape of the ancestral distribution may then be understood as follows. 
For $N=10000$, $s,\nu$ as in Fig. \ref{fig:stat_backandforth},
and $u$ well below $0.0005$, 
simulation according to the `recipe' resulting from \eqref{eq:a}
is similar to drawing from the  present distribution
until a fit individual is found, which is then identified as the
ancestor --- this is because $\lambda_j$ is close to 1 for $j \leq 100$;
and the expected frequency of fit individuals in the
present population, $p(\e_0)$, is not too small, so that, typically, only few draws
are required to find a fit individual. Therefore, the ancestors
are almost all fit, although the present population is not --- 
at the expense of an increasing number of virtuals. With
increasing $u$ (and, correspondingly, decreasing $p(\e_0)$), however,
one needs more and more trials to find a fit individual; this leads to
a further increase in the number of virtuals,
increasing $j$ and decreasing $\lambda_j$.
As a consequence, it becomes more and more likely 
to `accept' an unfit individual as the ancestor; finally 
(beyond $u \sim 0.0005$), both the present and the ancestral
population consist of mainly unfit individuals.  But a large number of
virtual branches is still present, because, for small $j$, the $\lambda_j$
are still close to one.
The number of virtual branches only decreases slowly, in accordance with
the decrease of the $\lambda_j$  with $u$.

\section{Acknowledgement.}
It is our  pleasure to thank Michael Baake,
Paul Fearnhead, Steve Krone, and Jay Taylor
for enlightening discussions and help with details, and
Jay Taylor and John Wakeley for critically reading the manuscript.
This work was supported by the Dutch-German Bilateral Research Group
on Random Spatial Models in Physics and Biology (DFG-FOR 498).

%\bibliography{../eb}

\newcommand{\noopsort}[1]{} \newcommand{\printfirst}[2]{#1}
  \newcommand{\singleletter}[1]{#1} \newcommand{\switchargs}[2]{#2#1}

\end{document}